\documentclass[a4paper]{jpconf}
\usepackage{graphicx}

\begin{document}

\title{Unified storage systems for distributed Tier-2 centres}

\author{G A Cowan$^1$, G A Stewart$^2$ and A Elwell$^2$}
\address{$^1$ Department of Physics, University of Edinburgh, Edinburgh, UK}
\address{$^2$ Department of Physics, University of Glasgow, Glasgow, UK}

\ead{g.cowan@ed.ac.uk}

\begin{abstract}
The start of data taking at the Large Hadron Collider will
herald a new era in data volumes and distributed processing in
particle physics. Data volumes of hundreds of Terabytes will be shipped
to Tier-2 centres for analysis by the LHC experiments using the
Worldwide LHC Computing Grid (WLCG).

In many countries Tier-2 centres are distributed between a number of
institutes, e.g., the geographically spread Tier-2s of GridPP in the
UK. This presents a number of challenges for experiments to utilise
these centres efficaciously, as CPU and storage resources may be
sub-divided and exposed in smaller units than the experiment would
ideally want to work with. In addition, unhelpful mismatches
between storage and CPU at the individual centres may be seen, which
make efficient exploitation of a Tier-2's resources difficult.

One method of addressing this is to unify the storage across a
distributed Tier-2, presenting the centres' aggregated storage as a
single system. This greatly simplifies data management for the
VO, which then can access a greater amount of data across the
Tier-2. However, such an approach will lead to scenarios where
analysis jobs on one site's batch system must access data hosted on
another site.

We investigate this situation using the Glasgow and Edinburgh
clusters, which are part of the ScotGrid distributed Tier-2. In
particular we look at how to mitigate the problems associated with
``distant'' data access and discuss the security implications of
having LAN access protocols traverse the WAN between centres.
\end{abstract}

\section{Introduction}

One of the key concepts behind Grid computing is the transparent
use of distributed compute and storage resources. Users of the Grid
should not need to know where their data resides, nor where it is
processed and analysed.

When the Large Hadron Collider at CERN begins to run at
luminosities sufficient for physics studies, it will produce around 15
petabytes of data a year. In order to analyse such a large quantity of
information, the Worldwide LHC Computing Grid (WLCG) has been
created. This is an international collaboration of physics laboratories and
institutes, spread across three major grids (EGEE, OSG and Nordugrid).

The UK's Grid for particle physics (GridPP) \cite{gridpp} started in
2001 with the aim of creating a computing grid that would meet the
needs of particle physicists working on the next generation of
particle physics experiments, such as the LHC. To meet this aim,
participating institutions were organised into a set of Tier-2 centres
according to their geographical location. ScotGrid \cite{scotgrid} is one such
distributed Tier-2 computing centre formed as a collaboration between the
Universities of Durham, Edinburgh and Glasgow.
To the Grid, the three collaborating institutes appear as individual sites.
Currently, the close association between them exists only at a managerial
and technical support level. One of the aims of this paper is to study the
possibility of having even closer-coupling between the sites of ScotGrid
(or other collections of sites on the Grid) from the point of view of
accessing storage resources from geographically distributed computing
centres.

Current estimates suggest that the maximum rate with which a physics analysis
job can read data is 2MB/s. We want to investigate if such a rate is possible
using distributed storage when using production quality hardware that is
currently operational on the WLCG grid. Physics analysis code will use the
POSIX-like LAN access protocols to read data, which for DPM is the Remote File
I/O protocol (RFIO).

Data transport using GridFTP across the WAN access has been considered
previously \cite{wan-access-paper}.

The structure of this paper is as follows. Section \ref{sec:middleware} describes the storage
middleware technology that we use in this study. We further
motivate this work in Section \ref{sec:wan}. Section \ref{sec:hardware}
discusses the storage, compute and networking hardware that is employed to
perform the testing. Section \ref{sec:testing} explains the reasoning behind our
testing methodology. We present and interpret the results of this testing in
Section \ref{sec:results}, present future work in \ref{sec:future} and conclude
in Section \ref{sec:conclusions}.

\section{Storage middleware\label{sec:middleware}}

The Disk Pool Manager (DPM) \cite{dpm} is a storage middleware product
created at CERN as part of the EGEE \cite{egee} project. It has
been developed as a lightweight solution for disk storage management at Tier-2
institutes. {\it A priori}, there is no limitation on the amount of disk space
that the DPM can handle.

\subsection{Architecture}\label{sec:arch}

The DPM consists of the following component servers,
\begin{itemize}
\item DPM ({\tt dpm}): keeps track of all requests for file access.
\item DPM name server ({\tt dpnsd}): handles the namespace for all files under the
control of DPM.
\item DRPM RFIO ({\tt rfiod}): handles the transfers for the RFIO protocol (See section
\ref{sec:protocols}).
\item DPM GridFTP ({\tt dpm-gsiftp}): handles data transfers requiring use of the
GridFTP protocol (See Section \ref{sec:protocols}).
\item Storage Resource Manager ({\tt srmv1}, {\tt srmv2}, {\tt srmv2.2}):
receives the SRM requests, passing them on to the DPM server.
\end{itemize}

\begin{figure}[t]
\centering
\includegraphics[scale=0.5]{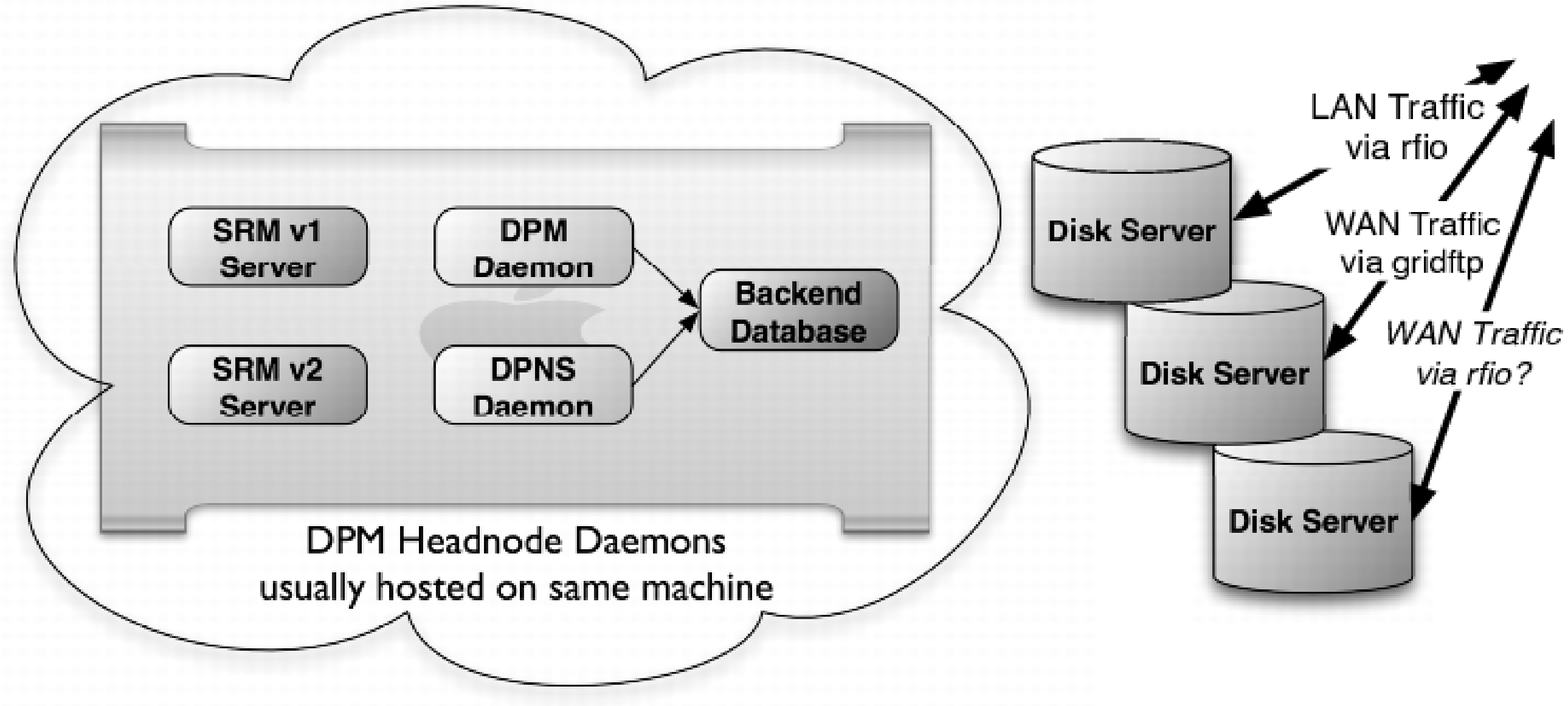}
\caption{Shows one possible configuration of the DPM server components.\label{fig:dpm_arch}}
\end{figure}

The protocols listed above will be described in the Section \ref{sec:protocols}.
Figure \ref{fig:dpm_arch} shows how the components can be configured in an
instance of DPM. Typically at a Tier-2 the server daemons
(\texttt{dpm}, \texttt{dpns}, \texttt{srm}) are shared on
one DPM \textit{headnode}, with separate large disk servers actually
storing and serving files, running \texttt{dpm-gsiftp} and \texttt{rfiod} servers.

\subsection{Protocols}\label{sec:protocols}

DPM currently uses two different protocols for data transfer and one for storage
management,

\begin{itemize}
\item GridFTP: typically used for wide area transfer of data files, e.g.,
movement of data from Tier-1 to Tier-2 storage.
\item Remote File IO (RFIO): GSI-enabled \cite{gsi} protocol which provides
POSIX \cite{posix} file operations, permitting byte-level access to files.
\item Storage Resource Manager: This standard interface is used on the WLCG grid
to permit the different storage server and client implementations to
interoperate.
\end{itemize}

RFIO is the protocol that physics analysis code should use in order to read data
stored within a DPM instance and is the protocol used to perform the tests in this
paper. Client applications can link against the RFIO client library permits
byte level access to files stored on a DPM. The library allows for four
different modes of operation,

\begin{itemize}
  \item {\tt 0}: Normal read with one request to the server.
  \item {\tt RFIO$\_$READBUF}: an internal buffer is allocated in the client API,
   each call to the server fills this buffer and the user buffer is filled
   from the internal buffer. There is one server call per buffer fill.
  \item {\tt RFIO$\_$READAHEAD}: {\tt RFIO$\_$READBUF} is  forced on and an
  internal buffer is allocated
  in the client API, Then an initial call is sent to the server which pushes data to
  the client until end of file is reached or an error occurs or a new request
  comes from the client.
  \item {\tt RFIO$\_$STREAM} (V3):This read mode opens 2 connections
     between the client and server, one data socket and one control
     socket. This allows the overlap of disk and network
     operations. Data is pushed on the data socket until EOF is
     reached. Transfer is interrupted by sending a packet on the
     control socket.
\end{itemize}

\section{RFIO over the WAN\label{sec:wan}}

Current Grid middleware is designed such that analysis jobs are sent to the site
where the data resides. The work presented in this paper presents an alternative
use case where analysis jobs can use RFIO for access to data held on a DPM which
is remote to the location analysis job is processed. This is of interest due to
a number of reasons,

\begin{itemize}
  \item Data at a site may be heavily subscribed by user analysis jobs, leading
  to many jobs being queued while remote computing resources remain under used.
  One solution (which is currently used in WLCG) is to replicate the data
  across multiple sites, putting it close to a variety of computing centres.
  Another would be to allow access to the data at a site from remote centres,
  which would help to optimise the use of Grid resources.
  \item The continued expansion of national and international low latency
  optical fibre networks suggest that accessing data across the wide area
  network could provide the dedicated bandwidth that physics analysis jobs will
  require in a production environment.
  \item Simplification of VO data management models due to the fact that any
  data is, in essence, available from any computing centre. The ATLAS computing
  model already has the concept of a ``cloud'' of sites which store datasets.
\end{itemize}

\subsection{Security}

RFIO uses the Grid Security Infrastructure (GSI) model, meaning that clients
using the protocol require X.509 Grid certificate signed by a trusted
Certificate Authority. Therefore, within the framework of x.509, RFIO can be
used over the wide area network without fear of data being compromised.

Additional ports must be opened in the site firewall to allow access to clients using
RFIO. They are listed below. Data transfer will use the site defined RFIO port
range.

\begin{itemize}
\item 5001: for access to the RFIO server.
\item 5010: for namespace operations via the DPNS server.
\item 5015: for access to the DPM server.
\end{itemize}

\section{Hardware setup\label{sec:hardware}}

\subsection{DPM server}

YAIM \cite{yaim} was used to install v1.6.5 of DPM on a dual core disk
server with 2GB of RAM. The server was running SL4.3 32bit with a 2.6.9-42
kernel. VDT1.2 was used \cite{vdt}. All DPM services were deployed on the same
server. A single disk pool was populated with a 300GB filesystem.

\subsection{Computing cluster}
To facilitate the testing, we had use of the UKI-SCOTGRID-GLASGOW WLCG grid site
\cite{scotgrid}. The computing cluster is composed of 140 dual core,
dual CPU Opteron 282 processing nodes with 8GB of RAM each. Being a production
site, the compute cluster was typically processing user analysis and
experimental Monte Carlo production jobs while our performance studies were
ongoing. However, observation showed that jobs on the cluster were typically CPU
bound, performing little I/O. As our test jobs are just the opposite (little
CPU, I/O and network bound) tests were able to be performed while the
cluster was still in production.\footnote{In fact this is quite
  reasonable, as in modern multicore SMP systems analysis jobs will share
  nodes with additional batch jobs, which will typically be CPU intensive.}

\subsection{Networking}

\begin{figure}[t]
\centering
\begin{tabular}{lr}
\includegraphics[scale=0.4]{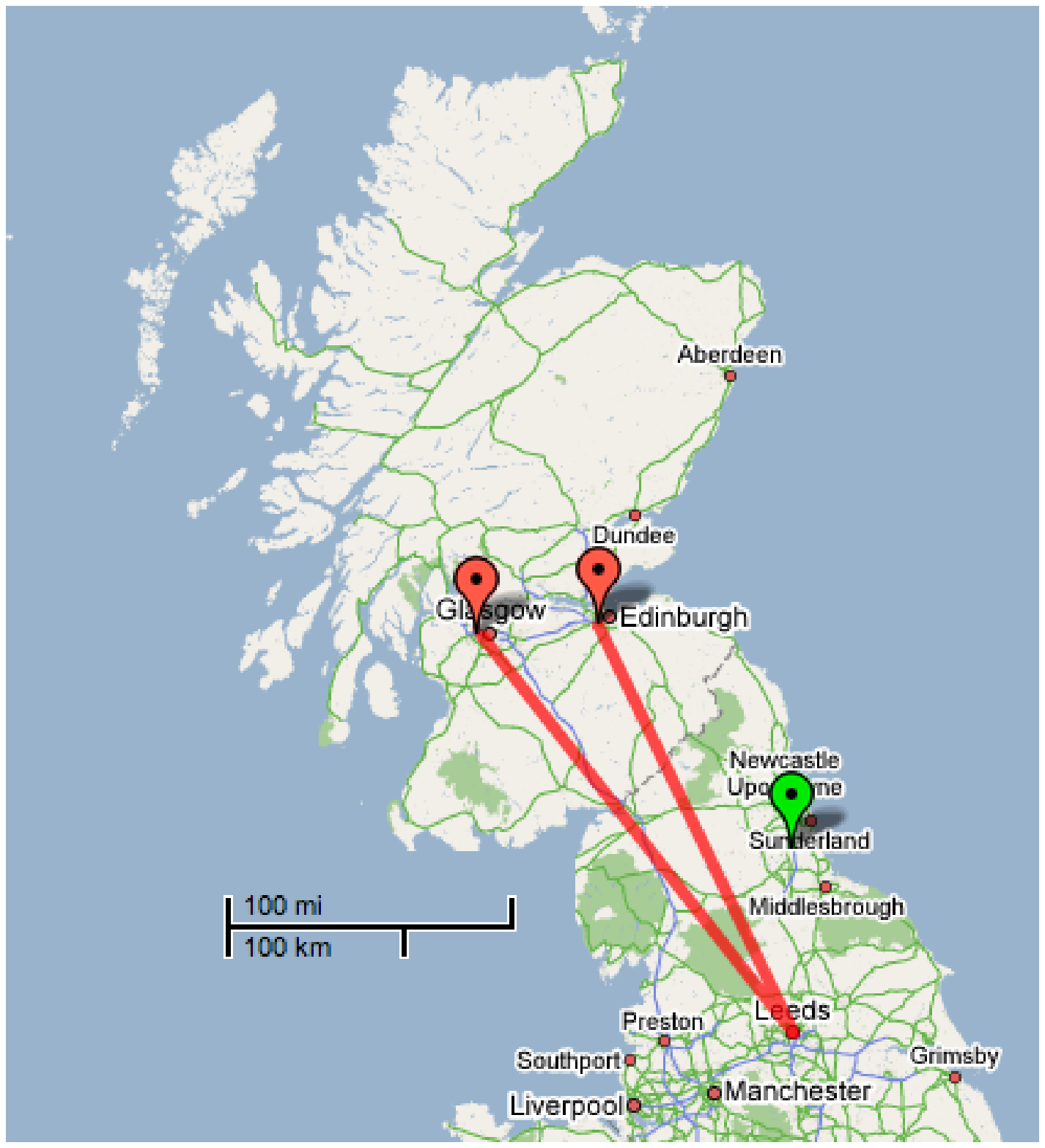}&
\hspace{1.5cm}\includegraphics[scale=0.4]{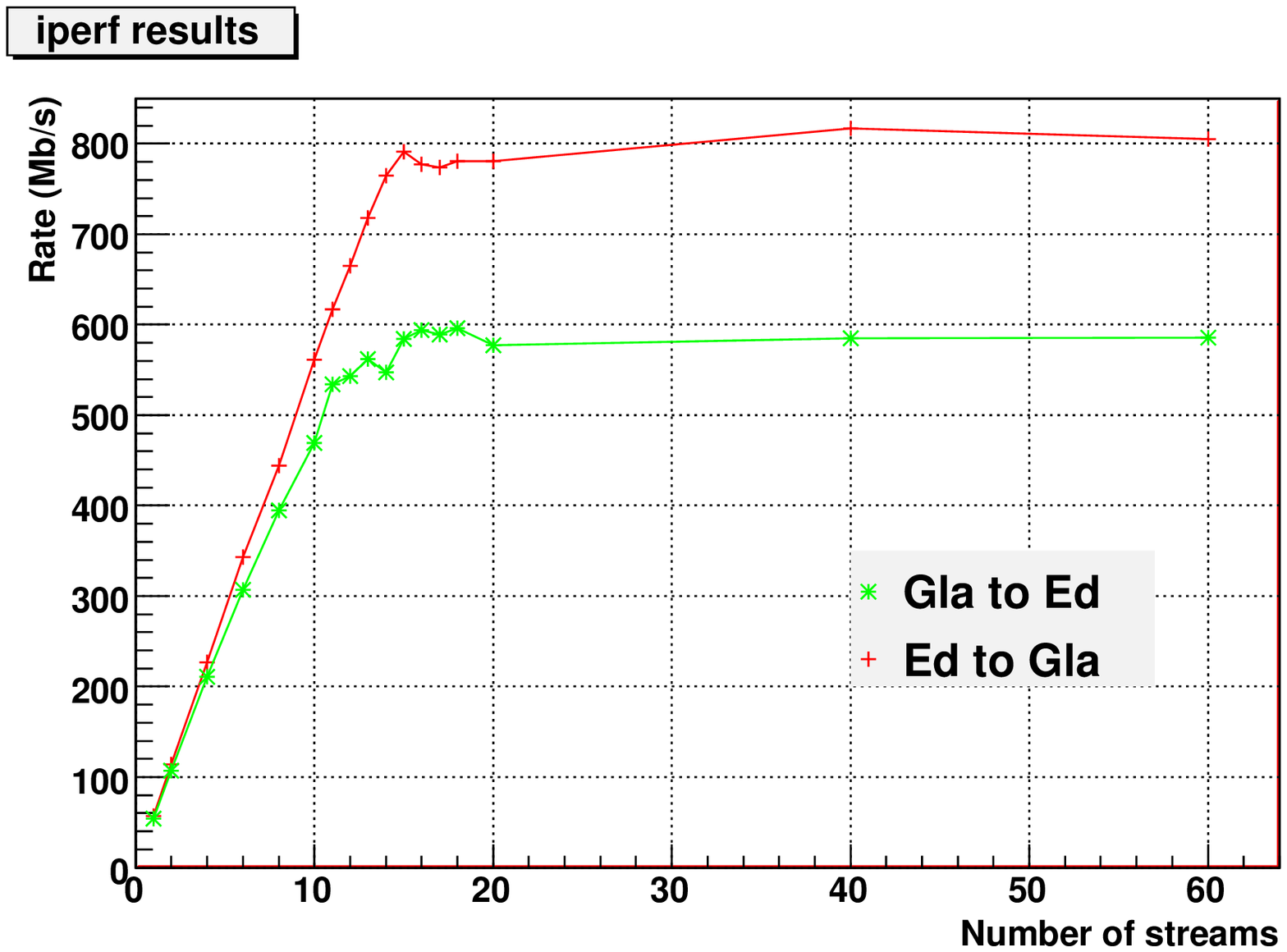}
\end{tabular}
\caption{{\bf Left:} The JANET-UK network path taken between the
UKI-SCOTGRID-GLASGOW
and ScotGRID-Edinburgh sites used in the RFIO testing. {\bf Right:} Iperf was
used to test the network between the two sites.}
\label{fig:path}
\end{figure}

The clients and servers at UKI-SCOTGRID-GLASGOW and ScotGRID-Edinburgh are
connected (via local campus routing) to the JANET-UK production academic network
\cite{janet} using gigabit ethernet. Figure \ref{fig:path} shows the results of
running iperf between the two sites.
This shows that the maximum file transfer rate that we could hope
to achieve in our studies is approximately 900Mb/s (100MB/s). The round trip
time for this connection is 12ms.

\begin{figure}[t]
\centering
\includegraphics[scale=0.5]{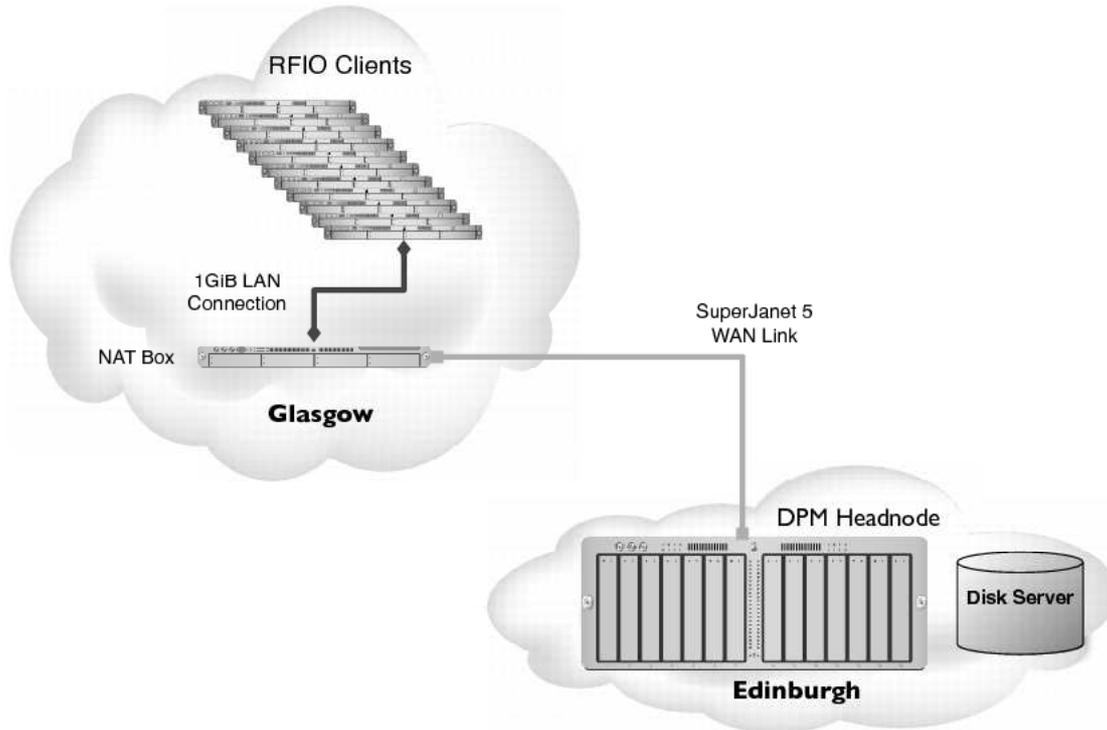}
\caption{The hardware setup used during the tests and the network connection
between sites.\label{fig:cluster}}
\end{figure}

\section{Methodology\label{sec:testing}}
\subsection{Phase space of interest}

Analysis jobs will read data in storage elements, so we restrict our
exploration to jobs which read data from DPM. We explore the effects
of using different RFIO reading modes, setting different RFIO buffer sizes and
client application block sizes. Since we are studying transfers across the WAN,
we also look at the effect of varying TCP window sizes on the total data
throughput.

\subsection{RFIO client application\label{client}}

In order to explore the parameter space outlined above, we developed our
own RFIO client application. Written in C, this application links against the
RFIO client library for DPM (\texttt{libdpm}). The application was designed such that it
can simulate different types of file access patterns. In our case, we
were interested in using the client where it sequentially reads blocks of a file
and also the case where it reads a block, skips ahead a defined number of blocks
and then reads again. This access was used to simulate the file access as used in
physics code as the job jumps to different parts of the file
when scanning for interesting physics events. Importantly, the client could be
configured to use one of the RFIO modes described in Section
\ref{sec:protocols}.

We did not use STREAM mode when skipping through the file as there appears to be
a problem with the pointer returned by the {\tt rfio\_seek} method such that it
does not report the correct position in the file.

\subsection{Source data and client initialisation}
\label{sec:source-data-init}

As, in general, one expects that each analysis job is uncorrelated with
the others, and so will be reading different data, 100 source data
files of 1GiB size were seeded onto the DPM. Each client then read a
single unique file. If this assumption were not valid, and several
clients were reading the same data, then the disk server would have
the opportunity to cache data from the file in question, possibly
resulting in faster reads.

We also choose to stress the DPM server itself during our testing
scenario by starting each of the clients within 1s. This should be
considered a worst case scenario for the storage system, as in
practice it is very unlikely that jobs will request opens in such
close succession.

Therefore, our testing scenario is deliberately setup in order
to stress the storage system and the network. This is essential in order to
establish whether or not RFIO access across the WAN can meet the highly
demanding data processing rates of LHC physics analysis jobs.



\section{Results\label{sec:results}}
\subsection{Client results}

The results presented below show client data transfer rates, defined as
\texttt{BYTES\_READ}/(\texttt{Open Time + Read Time}). It should be noted that
theoretical network bandwidth must also include IP and TCP overheads.
  
\subsubsection{Complete file reads\label{sec:complete}}

\begin{figure}[t]
\centering
\includegraphics[scale=0.5]{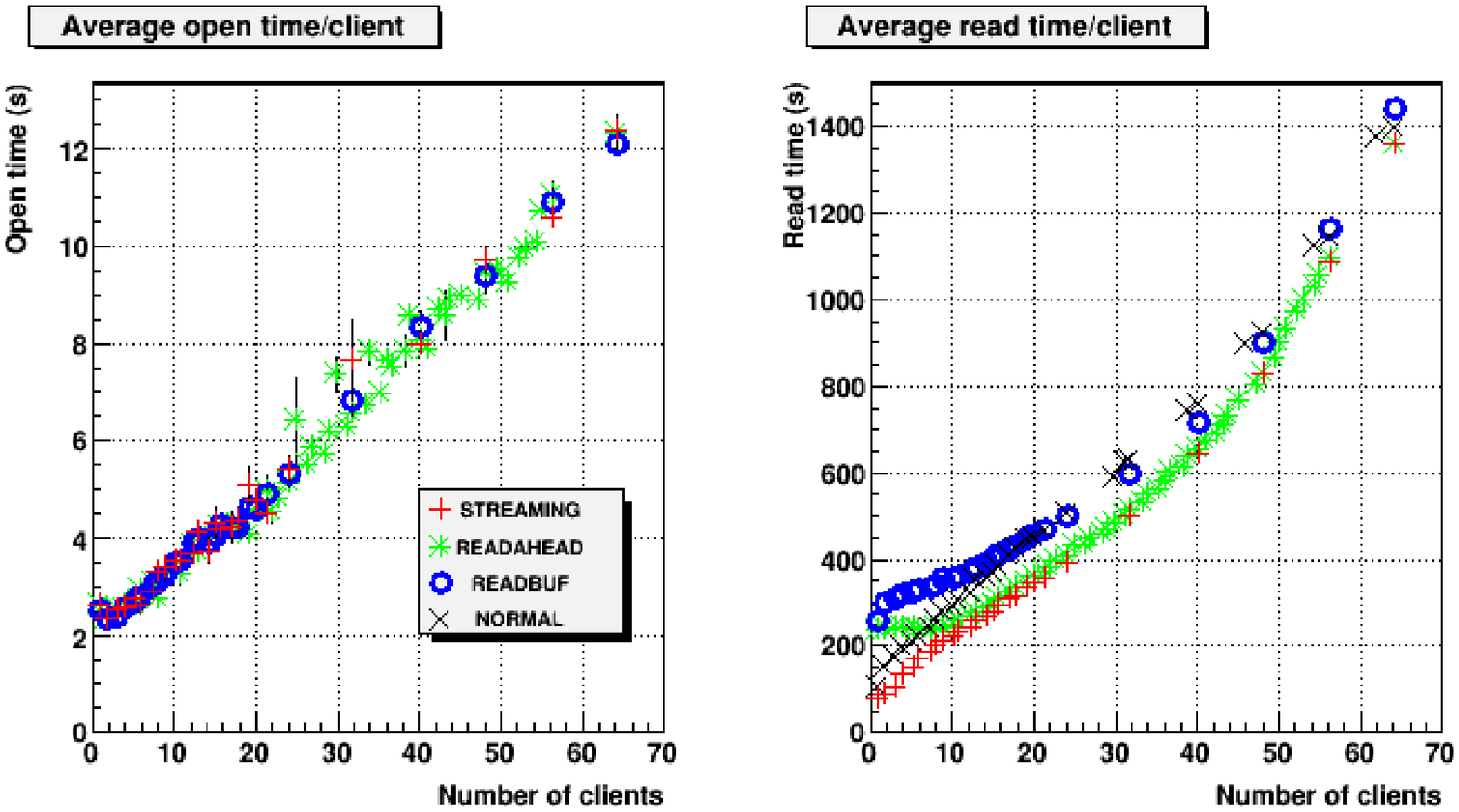}
\includegraphics[scale=0.5]{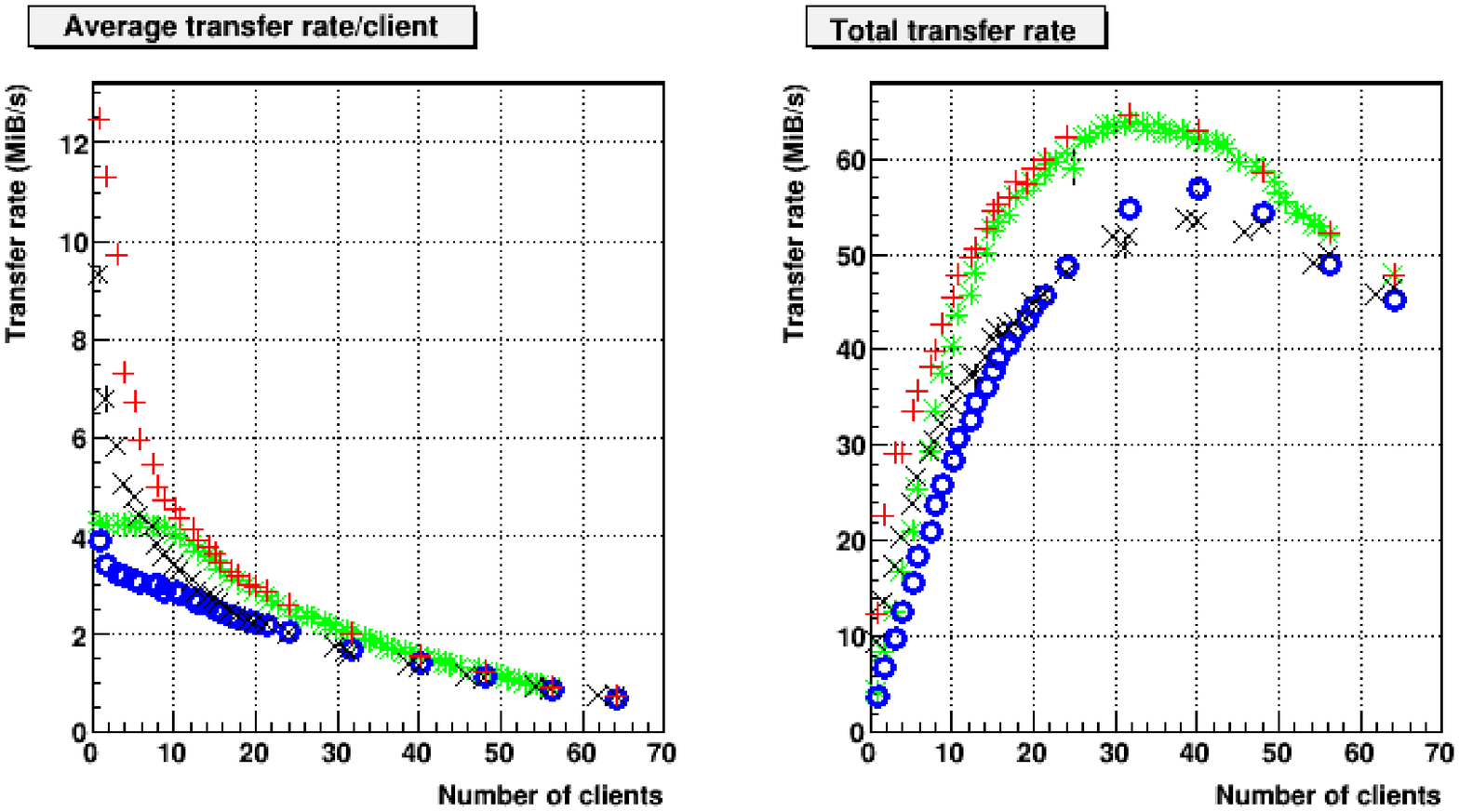}
\caption{Complete sequential reading of 1GiB files stored in the DPM from the
compute
cluster, for varying read modes and client number. The block size was 1MiB. Error
bars are smaller than the plotted symbols.}
\label{fig:seq-read-1048576-0-0-1-s0-n1}
\end{figure}

Figure \ref{fig:seq-read-1048576-0-0-1-s0-n1} shows the results for
reading 1GiB files. After only two or three simultaneous clients, the file open
time begins to increase approximately linearly with the number of clients, from
$\sim $2s up to $>$12s for 64 clients.

Reading rate is 4-12MiB/s for a single client, and rises rapidly for
small numbers of clients, increasing up to 62MiB/s for 32 clients using the
STREAMING and READAHEAD modes. These buffered modes perform better than READBUF
as both aggressively read data from the server. For larger client numbers the
transfer rates begin to tail off.This effect is caused by the disk server 
becoming I/O bound when serving data to so many clients at once, with the
bottleneck of many head movements on the disk to jump between the many open
files.

\begin{figure}[t]
\centering
\includegraphics[scale=0.5]{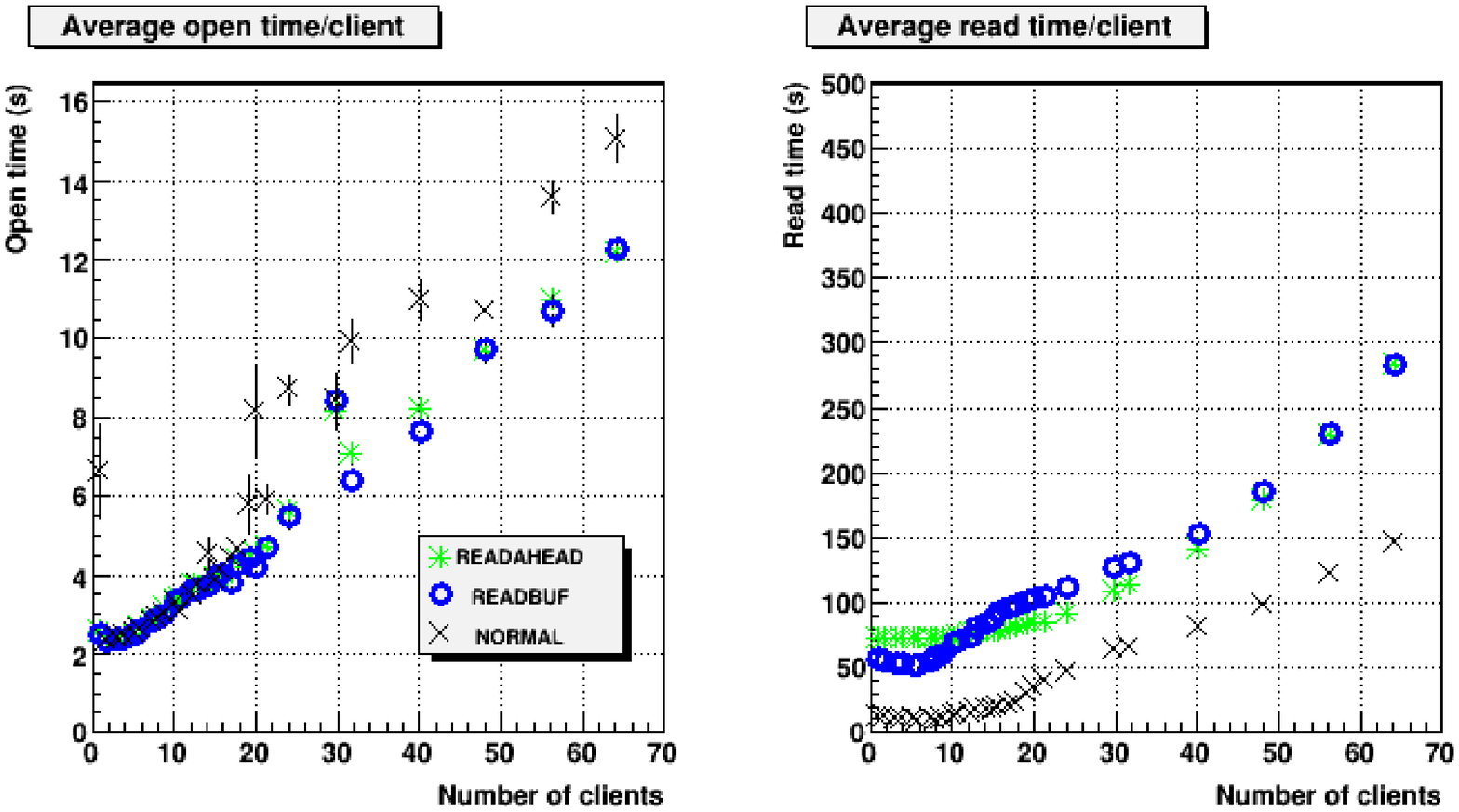}
\includegraphics[scale=0.5]{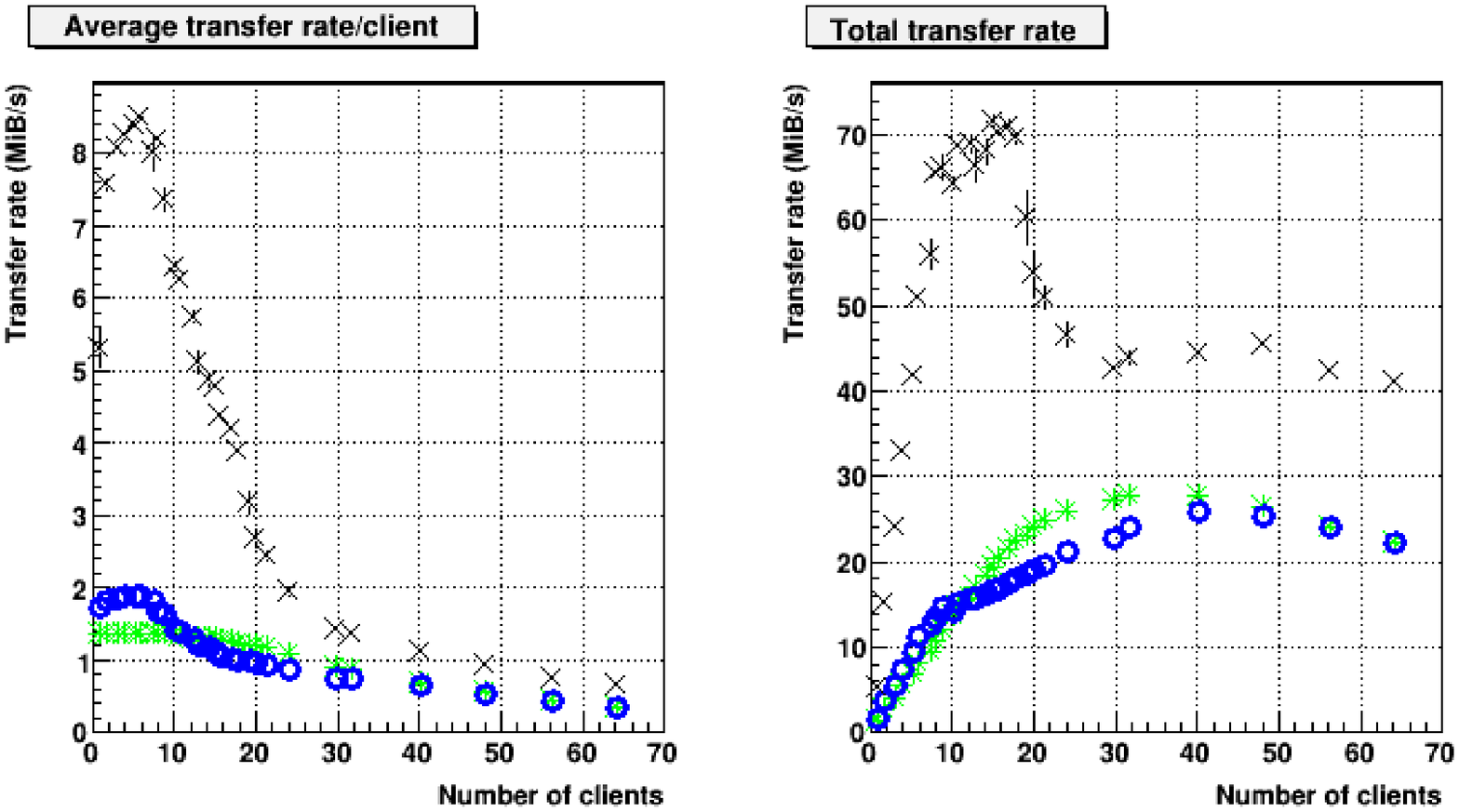}
\caption{Results for partial (10\%) reading of 1GiB files stored in the DPM from
the Glasgow compute cluster for the different read modes and varying number of
simultaneous clients. The block size was 1MiB. The black lines show the RMS errors.}
\label{fig:ran-read-1048576-0-0-1-s9-n1}
\end{figure}

\subsubsection{Partial file reads}
\label{sec:partial-file-reads}

Results for partial file reads are shown in Figure
\ref{fig:ran-read-1048576-0-0-1-s9-n1}. In this case each
client reads 1MiB from the source file, then skips 9MiB (simulating
reading 10\% of the events in, say, an AOD file). As expected the
results for opening files are very similar to Figure
\ref{fig:seq-read-1048576-0-0-1-s0-n1} -- the \texttt{rfio\_open()}
call is exactly same as the previous case.

Read rates, as expected, for READBUF and READAHEAD modes are considerably lower
than for complete reads, as the disk server has to skip large portions of the
file, repositioning the reading heads. Maximum rates in this case are only
2MiB/s. In stark contrast the case when complete files are read,
NORMAL mode performs better than the buffer reading modes, particularly at small
client number. In particular, the total transfer rate for between 11 and 18
clients is larger than the maximum rate seen in Section \ref{sec:complete}.
It is not clear why this rate is not sustained beyond 18 clients; further
investigation is required. The advantage of NORMAL mode when skipping through
the file can be understood as the buffered modes read data which is not needed
by the client. This data is thus discarded, but has loaded both the disk server
and the network, reducing performance.

\subsubsection{RFIO IOBUFSIZE}

The value of the internal API buffer used by clients in the
default READBUF mode is set by the site administrator in
\texttt{/etc/shift.conf}, rather than by clients.

\begin{figure}[t]
\centering
\includegraphics[scale=0.42]{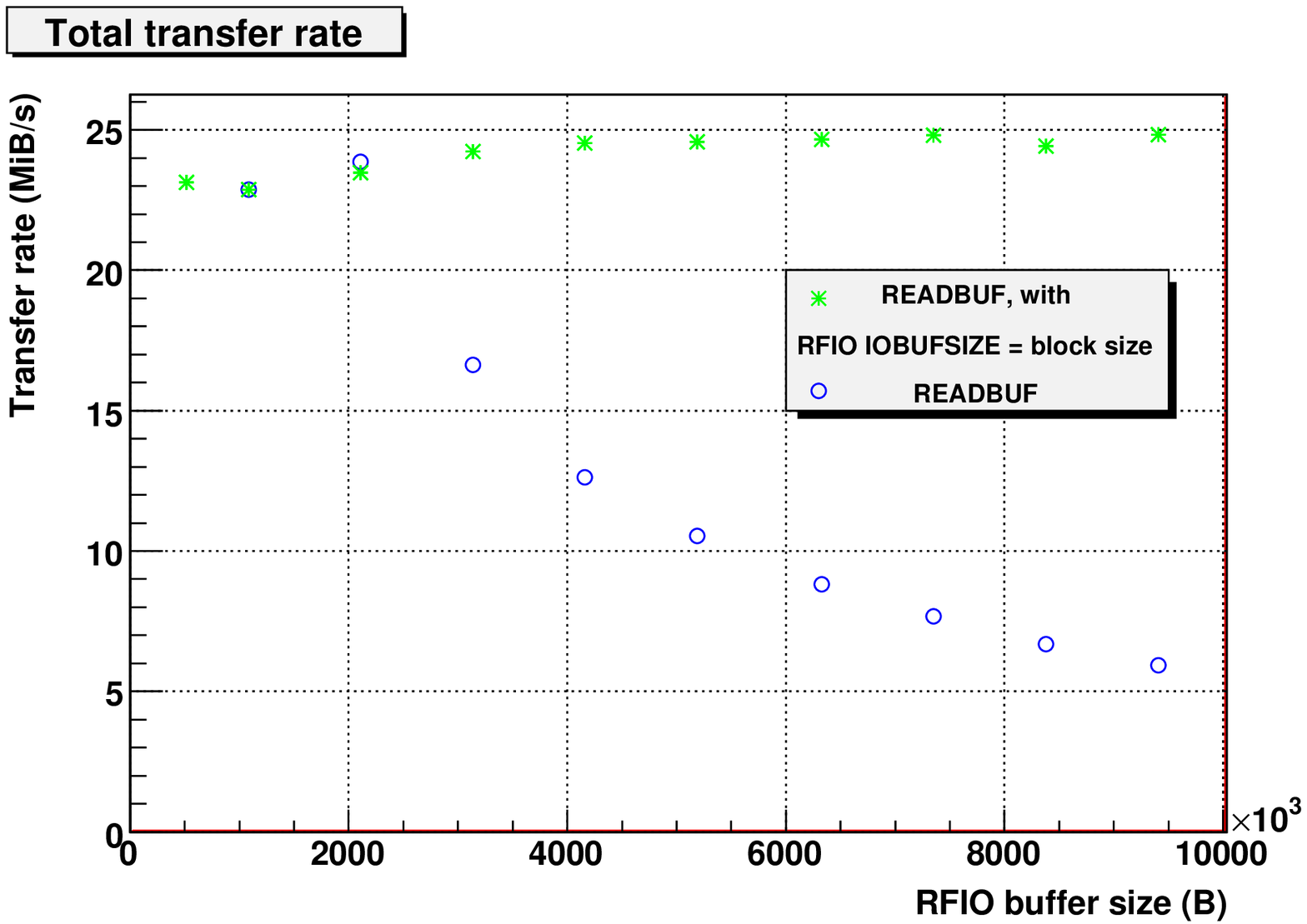}
\caption{Reading 1GB files stored in the DPM using RFIO READBUF mode from 30
client nodes. Transfer rate is plotted versus the RFIO IOBUFSIZE. Blue circles
show the case for partial file reads with a block size of 1MB while read pluses
show whole file reads with the block size set equal to the RFIO IOBUFSIZE.}
\label{fig:rfio-wan-cli30-buf1048576-s9-n1}
\end{figure}

Figure \ref{fig:rfio-wan-cli30-buf1048576-s9-n1} shows the results of the
variation of the total transfer rate with IOBUFSIZE when using READBUF mode to
read 10\% of the file. For partial file reads increasing the value of IOBUFSIZE
clearly hurts the overall rate considerably. This is caused by the internal
buffer being filled before the client actually requests data. In the case of
skipping through a file the client in fact does not require this data
and so network and I/O bandwidth has been consumed needlessly. For the case
where the client application block size is altered so as to match
that of the IOBUFSIZE, there is essentially no change in the total transfer rate
as the IOBUFSIZE matches. This makes sense as the client is requesting the same
amount of data as is being filled in the RFIO buffer, meaning bandwidth is not wasted.

This study has shown that the value of RFIO IOBUFSIZE should be left at its
default setting of 128kB. In particular setting too high a value will penalise
clients using the default READBUF mode to make partial reads of a file.

\subsubsection{TCP tuning}

Since we are moving data across the WAN, we decided to study the effect of TCP
window sizes on the throughput of our tests. We modified the
{\tt /etc/syscont.conf} settings on the client side in the following way,\\

\small
\hspace{-0.6cm}
\begin{tabular}{llllll}
{\tt net.ipv4.tcp$\_$rmem = 4096 \$TCPVALUE \$TCPVALUE}&&&&&{\tt net.core.wmem$\_$default = 1048576}\\
{\tt net.ipv4.tcp$\_$wmem = 4096 \$TCPVALUE \$TCPVALUE}&&&&&{\tt net.core.rmem$\_$max = \$TCPVALUE}\\
{\tt net.ipv4.tcp$\_$mem = 131072 1048576 \$TCPVALUE}&&&&&{\tt net.core.wmem$\_$max = \$TCPVALUE}\\
{\tt net.core.rmem$\_$default = 1048576}&&&&&
\vspace{0.2cm}\end{tabular}

\normalsize
Where {\tt \$TCPVALUE} $\in (0.5, 1, 2, 4, 8,16)$MiB. {\tt sysctl -p} was executed
after setting the window sizes. The default value used in all other tests
presented here was 1MiB.

\begin{figure}[t]
\centering
\includegraphics[scale=0.42]{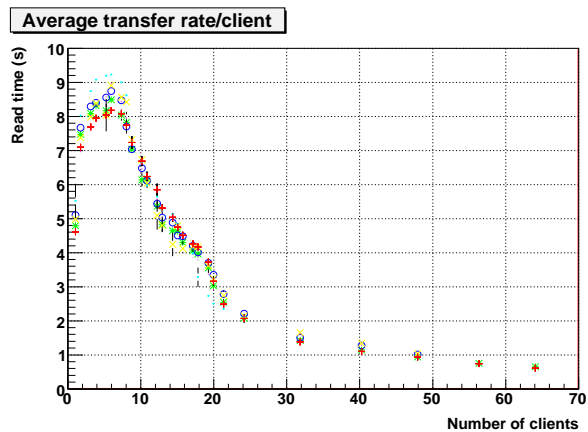}
\caption{
Read time and transfer rate as a function of client number for
different values of the TCP ``tuning" using the NORMAL RFIO mode when skipping
through the files. In all cases here, the value of the TCP ``tuning'' was set to
the same value as the client application block size and the RFIO buffer. The
results are essentially the same for all cases of the TCP tuning.}
\label{fig:rfio-wan-tcp-normal-s9-n1-common-tcp-buf-block}
\end{figure}

Figure \ref{fig:rfio-wan-tcp-normal-s9-n1-common-tcp-buf-block} shows how the
total transfer rate varies with client number as we alter the TCP tuning
parameters. In this case, we were skipping though 10\% of the file and setting
the tuning to the same value as the client application block size and the RFIO
buffer size. Different colours correspond to different TCP window sizes in the
range specified above. It is clear that there is very little variation in the
total observed rate as a function of window size. This is
to be expected when such a large number of clients are simultaneously reading as
they each have only a small part of the total bandwidth available to them. A
small improvement is seen at small client numbers with a larger window (the
turquoise points). It is likely that additional improvements in the transfer
rate will only come after optimisations have been made in the client
application.

\subsection{Comparison with LAN access\label{sec:lan}}

\begin{figure}[t]
\centering
\begin{tabular}{cc}
\includegraphics[scale=0.4]{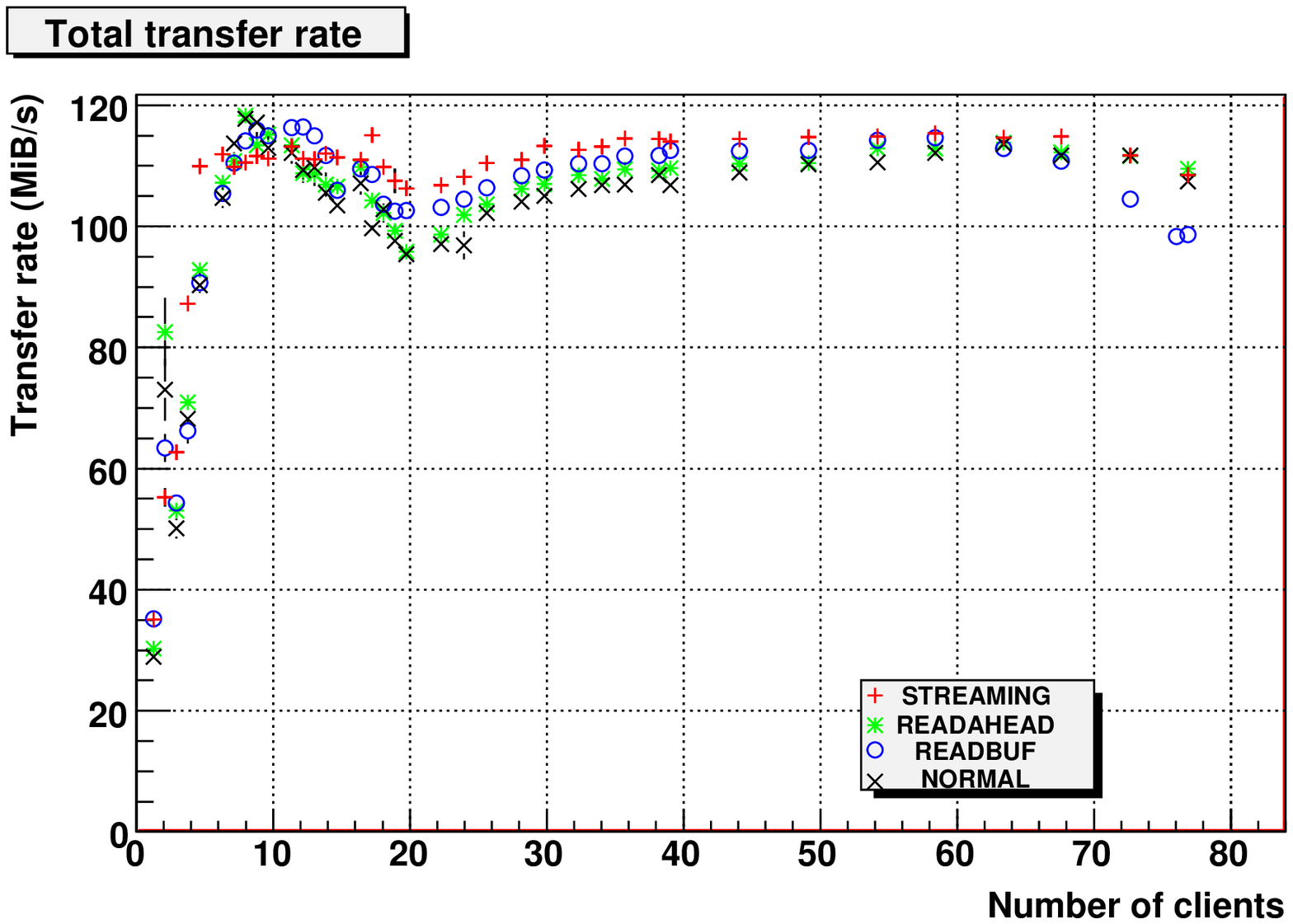}&
\includegraphics[scale=0.4]{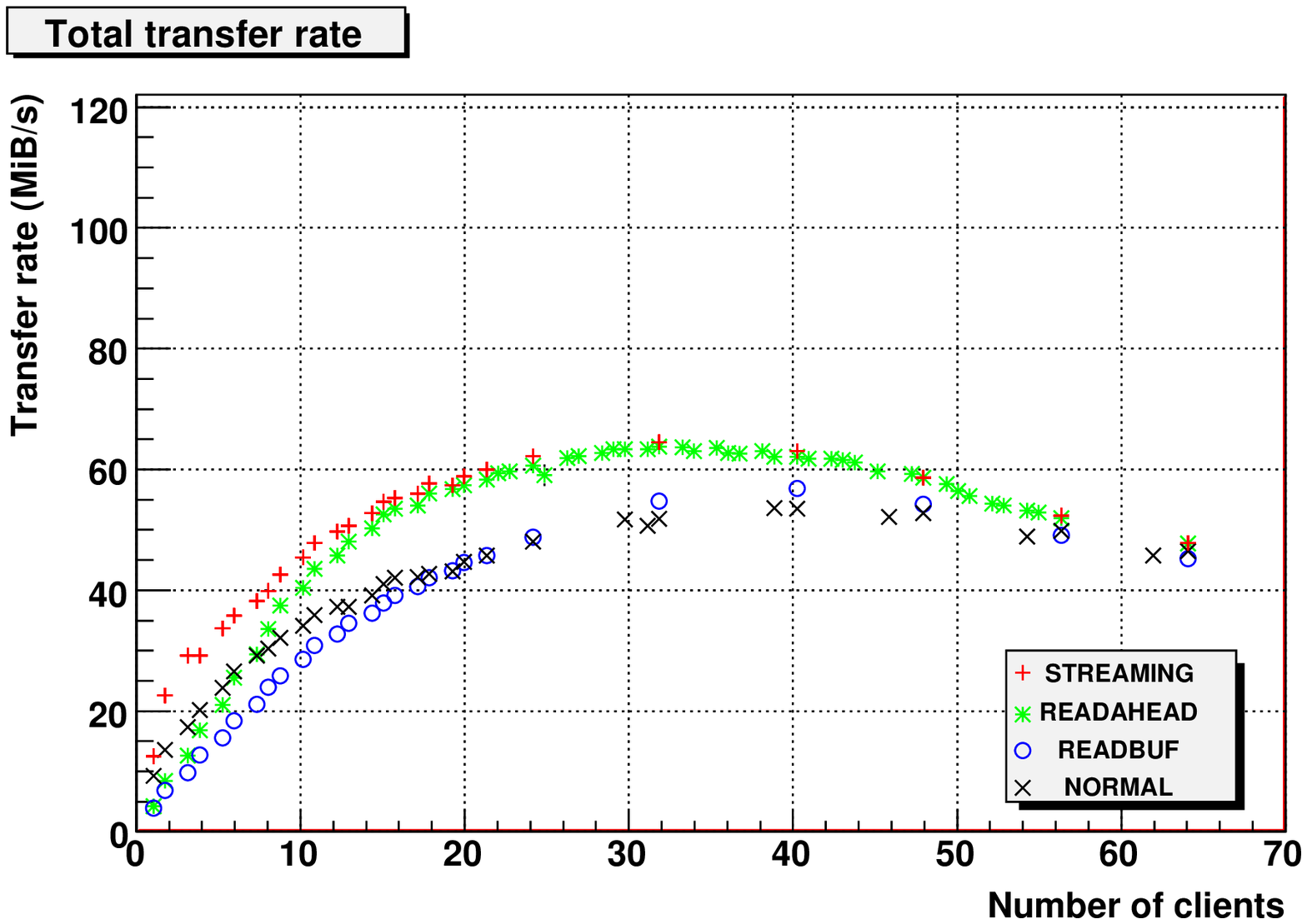}
\end{tabular}
\caption{{\bf Left:} Shows the total rate when reading data
sequentially from the local Glasgow DPM to clients on the cluster, across the LAN.
{\bf Right:} Shows the equivalent data access when reading data
from the Edinburgh DPM across the WAN (repeated from Figure
\ref{fig:seq-read-1048576-0-0-1-s0-n1} for clarity).}
\label{fig:lan}
\end{figure}

Figure \ref{fig:lan} shows a comparison of the total data rate as a function of
client number when reading files sequentially from a {\it single} DPM disk server
across the LAN and across the WAN. The LAN in this case is the local network of
the UKI-SCOTGRID-GLASGOW site which consists of a Nortel 5510 switch stack with
each cluster node connected through a 1GiB ethernet connection. This limits the
rate for each client to 1GiB, but more importantly limits the rate per-disk
server to 1GiB. Comparable hardware for the disk servers was used in each case.

Unsurprisingly, the total transfer rate across the dedicated LAN is larger than
that across the WAN, where we would expect a maximum throughput of around
100MiB/s (Figure \ref{fig:path}) as we are in contention for network resources
with
other users. However, we show that it is relatively simple to achieve reasonable
data rates in relation to LAN access. It will be interesting to study what 
throughput can be achieved across the WAN as the number of disk servers is increased. 

\subsection{Server results\label{sec:server}}
\subsubsection{RFIO open times and errors}
\label{sec:rfio-open-times}

\begin{figure}[t]
\centering
\begin{tabular}{cc}
\includegraphics[scale=0.4]{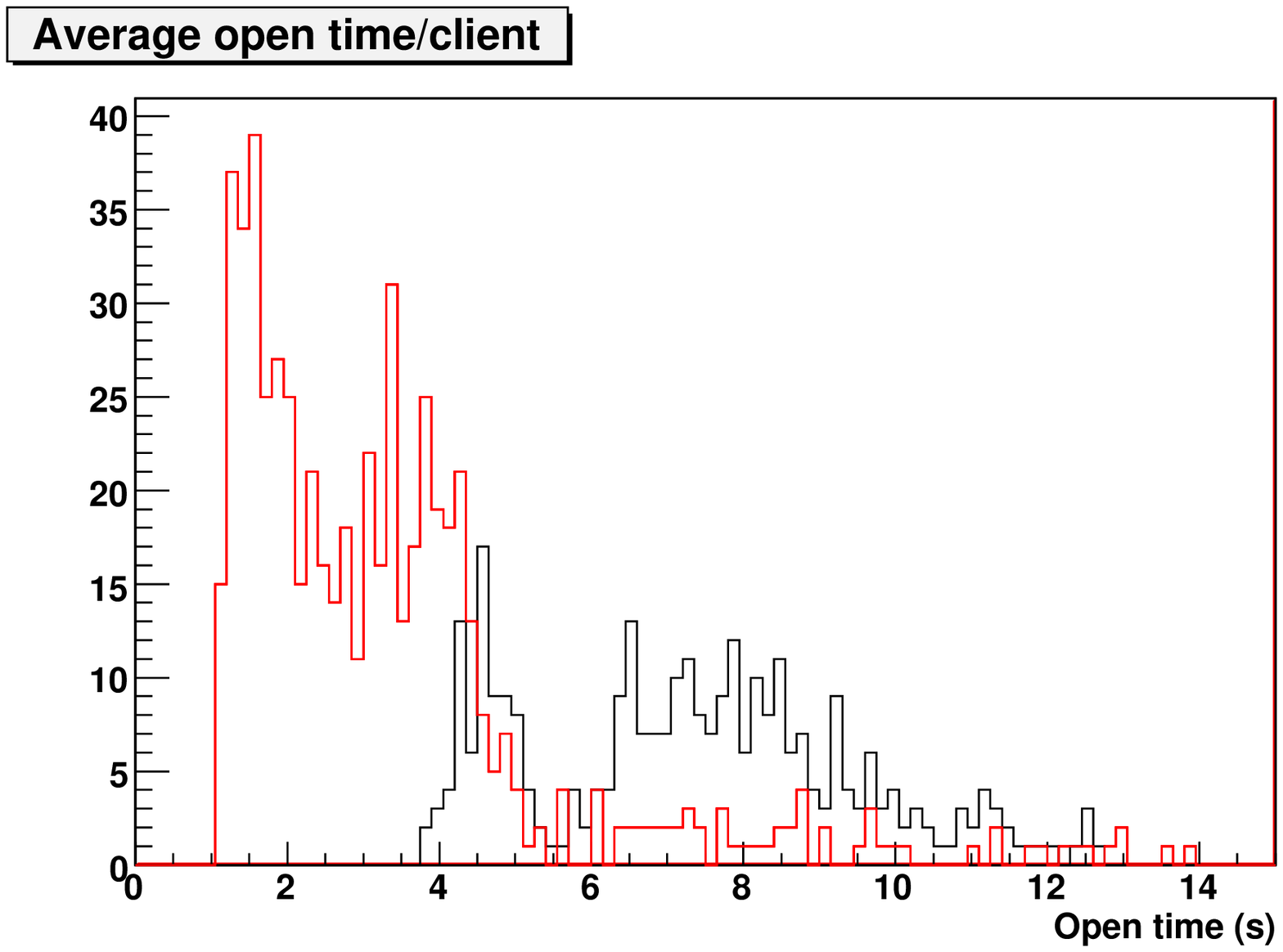}
&
\includegraphics[scale=0.4]{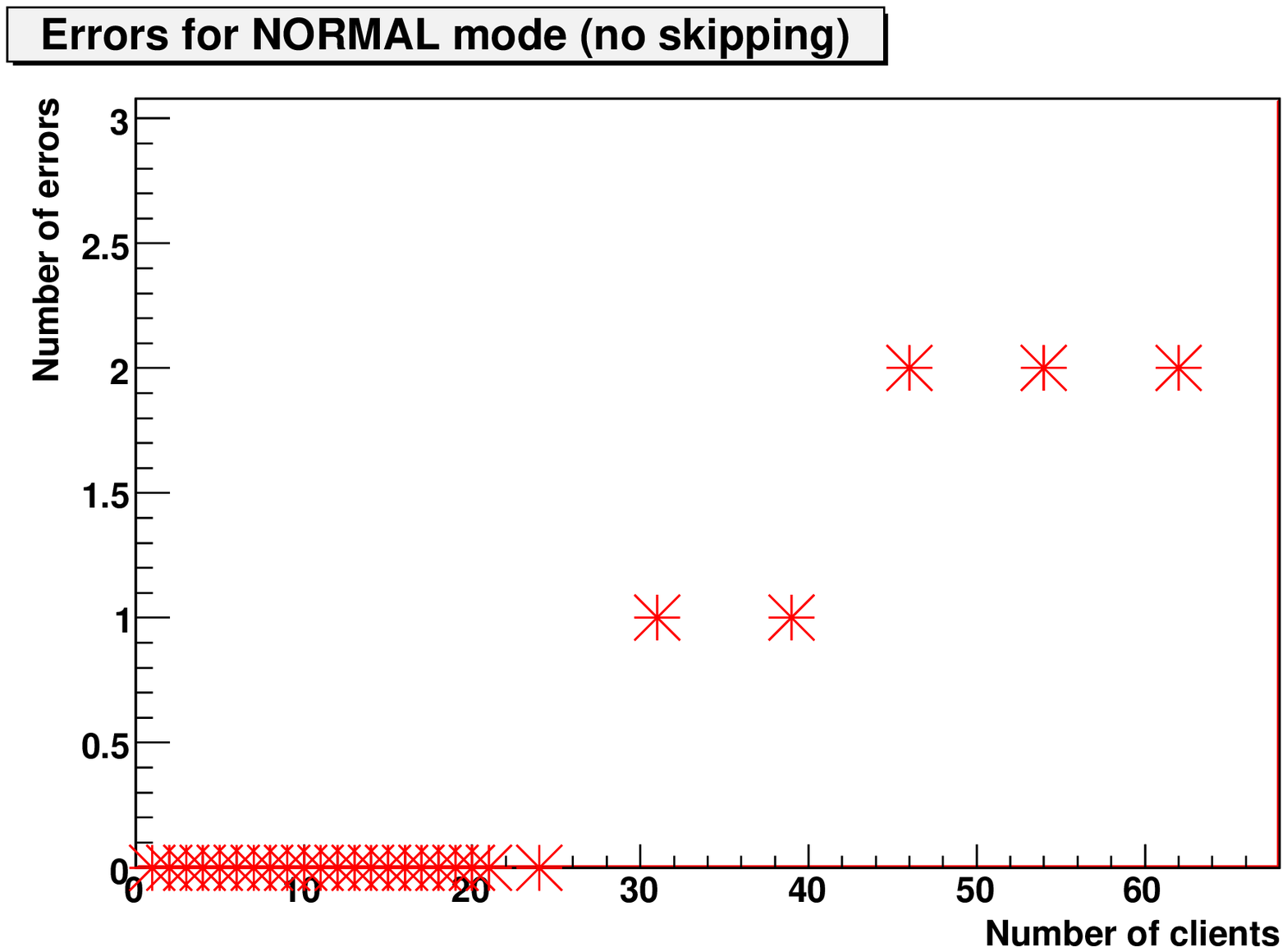}
\end{tabular}
\caption{{\bf Left:} Distribution of the open times for the different RFIO modes
(not
NORMAL). The red histogram is for $<20$ simultaneous clients, while the black
histograms is for $\ge 20$. {\bf Right:} File open errors, when multiple clients
attempt to open files.}
\label{fig:rfio-wan-openTime-histo-c20-no-normal}
\end{figure}

Figure \ref{fig:rfio-wan-openTime-histo-c20-no-normal} (Left) shows how the
average open time increases from 2 seconds for a small number of clients ($<
20$) up to around 8 seconds for a larger number of clients ($\geq 20$). Clearly,
this results is dependent on the hardware used during the testing. We collected
fewer statistics for client numbers $\ge 23$ due to the long run time of the tests.

While server performance clearly degrades when many clients
simultaneously attempt to open files, most files are opened
successfully, as can be seen in Figure
\ref{fig:rfio-wan-openTime-histo-c20-no-normal} (Right). This is a
substantial improvement over earlier versions of DPM, which could only
support about 40 opens/s \cite{dpm-jan-tests}.

%

\section{Future work\label{sec:future}}

\subsection{Networking}

As shown in Section \ref{sec:lan}, use of the production network for the data
transfer limited the data throughput relative to that of the LAN. We plan to
continue this work by making use of a newly provisioned lightpath between the
two Glasgow and Edinburgh. This will provide dedicated bandwidth with an RTT of around
2ms.

\subsection{Alternative tests}

We would like to study more realistic use cases involving real physics analysis
code. In particular, we would like to make use of the ROOT \cite{root} 
TTreeCache object which has been shown \cite{rfio-cache-test} to give efficient
access to ROOT objects across the WAN. As ROOT is the primary tool for analysing
physics data, it is essential that the performance benefits of this access
method are understood.

The tests performed so far have been designed to simulate access to the DPM by
clients that operate in an expected manner (i.e., open file, read some
data, close file). It would be an interesting exercise to perform a quantitative
study of the storage element and network when presented unexpected non-ideal
use cases of the protocol. 

\subsection{Distributed DPM}

The DPM architecture allows for the core services and disk servers (Section
\ref{sec:arch}) to be spread across different network domains. Therefore, rather
than having separate DPM instances at each site, we could create a single
instance that spans all collaborating institutes. However, there are
disadvantages to this this approach, as an inter-site network outage could
take down the entire system. Finally, DPM does not currently have the concept
of how ``expensive'' a particular data movement operation would be, which could
impact the behaviour and performance of this setup.


\subsection{Alternative protocols}

In addition to RFIO, xrootd has recently been added as an access protocol.
This implementation currently lacks GSI security \cite{gsi}, making it
unsuitable for use across the WAN. Once security is enabled, it would be
interesting to perform a similar set of tests to those presented above. Similar
tests should be performed when DPM implements v4.1 of the NFS protocol
\cite{nfs4.1}. 

\section{Conclusions\label{sec:conclusions}}

Through this work we have shown that it is possible to use RFIO to provide byte
level access to files stored in an instance of DPM across the wide area network.
This has shown the possibility of unifying storage resources across
distributed Grid computing and data centres, which is of particular
relevance to the model of distributed Tier-2 sites found within the UK GridPP
project.
Furthermore, this should be of interest to the data management operations of
virtual organisations using Grid infrastructure as it could lead to optimised
access to compute and data resources, possibly leading to a simpler data
management model.

Using a custom client application, we have studied the behaviour of RFIO access
across the WAN as a function of number of simultaneous clients accessing the
DPM; the different RFIO modes; the application block size and the
RFIO buffer size on the client side. We looked at the effect of varying the TCP
window size on the data throughput rates and found that it had little effect,
particularly when a large number of clients were simultaneously reading data.
Further work should be done to explore application optimisations before looking
at the networking stack.

Our testing has shown that RFIO STREAMING mode leads to the highest overall
data transfer rates when sequentially reading data. The
rates achieved were of order 62MiB/s on the production JANET-UK network between
the UKI-SCOTGRID-GLASGOW and ScotGRID-Edinburgh sites. For the case
where the client only accesses 10\% of the file, RFIO mode in NORMAL mode was
shown to lead to the best overall throughput as it does not transfer data that
is not requested by the client.

For all RFIO modes, file open times increase linearly with the number of
simultaneous clients, from $\sim 2$s with small number of clients up to $\sim
12$s with 64 clients. This increase is to be expected, but it is unclear at this
time how it will impact on actual VO analysis code. 

Finally, we have shown that it is possible to access remote data using a
protocol that is typically only used for access to local grid storage. This
could lead to a new way of looking at storage resources on the Grid and could
ultimately impact on how data is efficiently and optimally managed on the Grid.

\ack

The authors would like to thank all those who helped in the preparation of this
work. Particular mention should go to DPM developers for very helpful
discussions and B Garrett at the University of Edinburgh. This work was funded
by STFC/PPARC via the GridPP project. G Stewart is funded by the EU EGEE
project.


\section*{References}

\begin{thebibliography}{99}
\raggedright
\bibitem{gridpp}{\it http://www.gridpp.ac.uk}
\bibitem{scotgrid}{\it http://www.scotgrid.ac.uk/}
\bibitem{wan-access-paper}{Greig Cowan, Graeme Stewart, Jamie Ferguson. \textit{Optimisation of Grid Enabled Storage at Small Sites.} Proceedings of 6th UK eScience All Hands Meeting, Paper Number 664, 2006.}
\bibitem{dpm}{\it http://twiki.cern.ch/twiki/bin/view/LCG/DpmAdminGuide}
\bibitem{egee}{\it http://www.eu-egee.org}
\bibitem{gsi}{\it http://www.globus.org/toolkit/docs/4.0/security/}
\bibitem{posix}{\it http://standards.ieee.org/regauth/posix/index.html}
\bibitem{yaim}{\it http://twiki.cern.ch/twiki/bin/view/EGEE/YAIM}
\bibitem{vdt}{\it http://vdt.cs.wisc.edu/}
\bibitem{janet}{\it http://www.ja.net/}
\bibitem{dpm-jan-tests}{Graeme Stewart and Greig Cowan. \textit{rfio
      tests of DPM at Glasgow} WLCG Workshop, CERN January 2007.
    http://indico.cern.ch/contributionDisplay.py?contribId=101\&sessionId=13\&confId=3738.}
\bibitem{root}{\it http://root.cern.ch}
\bibitem{rfio-cache-test}{Rene Brun, Leandro Franco, Fons
  Rademakers. \textit{Efficient Access to Remote Data in High Energy
    Physics} CHEP07, Victoria, 2007. http://indico.cern.ch/contributionDisplay.py?contribId=284\&sessionId=31\&confId=3580.}
\bibitem{nfs4.1}{\it http://opensolaris.org/os/project/nfsv41/}
\end{thebibliography}
\end{document}